\documentclass[fleqn,usenatbib]{mnras}

\usepackage{graphicx}%
\usepackage{dcolumn}%
\usepackage{bm}%
\usepackage{graphics, caption}%
\usepackage[utf8]{inputenc}
\usepackage[T1]{fontenc}
\usepackage{amsmath,booktabs,braket,cleveref,mathptmx,xcolor,siunitx}
\usepackage{amssymb}	%
\usepackage{physics}
\usepackage{rotating}
\usepackage{siunitx}

\usepackage[version=4]{mhchem} 
\usepackage[normalem]{ulem}
\usepackage{array}

\newcommand{\abinitio}{\emph{ab initio}}
\newcommand{\MARVEL}{{\sc Marvel}}
\newcommand{\Marvel}{{\sc Marvel}}

\newcommand{\Duo}{{\sc Duo}}

\newcommand{\Mollist}{{\sc Mollist}}
\newcommand{\MOLLIST}{{\sc Mollist}}

\newcommand{\cm}{cm$^{-1}$}

\newcommand{\LLname}{Trihybrid}

\def\a0{{$a_{\rm 0}$}}

\newcommand{\alert}[1]{\textcolor{black}{ #1}}

\newcommand{\mc}{\multicolumn}
\newcolumntype{H}{>{\setbox0=\hbox\bgroup}c<{\egroup}@{}}

\newcolumntype{d}{D{.}{.}{-1}}

\usepackage{graphicx} %
\usepackage[T1]{fontenc}
\usepackage{lmodern}
\usepackage[english]{babel}
\usepackage[autostyle]{csquotes}
\usepackage[]{physics}
\usepackage{xcolor}
\usepackage{caption}
\usepackage{subfig}
\usepackage{siunitx}
\usepackage{booktabs}
\usepackage{multirow}
\usepackage{cleveref}
\usepackage{tabularx}
\usepackage[normalem]{ulem}
\useunder{\uline}{\ul}{}

\newcommand{\X}{X~${}^2\Sigma^+$}
\newcommand{\A}{A~${}^2\Pi$}
\newcommand{\B}{B~${}^2\Sigma^+$}

\newcommand{\asig}{a~${}^4\Sigma^+$}

\title{Full Spectroscopic Model and Trihybrid Experimental-Perturbative-Variational Line List for CN}%

\author[Syme and McKemmish]{
Anna-Maree Syme$^{1}$, Laura K. McKemmish$^{1}$\thanks{E-mail: l.mckemmish@unsw.edu.au}
\\
$^{1}$School of Chemistry, University of New South Wales, 2052, Sydney, Australia
}
\date{Accepted XXX. Received YYY; in original form ZZZ}

\pubyear{2021}

\begin{document}

\date{\today}

\maketitle

\begin{abstract}
Accurate line lists are important for the description of the spectroscopic nature of small molecules. While a line list for CN (an important molecule for chemistry and astrophysics) exists, no underlying energy spectroscopic model has been published, which is required to consider the sensitivity of transitions to a variation of the proton-to-electron mass ratio. 

Here we have developed a \Duo{} energy spectroscopic model as well as a novel hybrid style line list for CN and its isotopologues, combining energy levels that are derived experimentally (\Marvel{}), using the traditional/perturbative approach (\Mollist), and the variational approach (from a \Duo{} spectroscopic model using standard
ExoMol methodology). The final \LLname{} ExoMol-style line list for \ce{^{12}C^{14}N} consists of 28,004 energy levels (6,864 experimental, 1,574 perturbative, the rest variational) and 2,285,103 transitions up to 60,000 \cm{} between the three lowest electronic states (\X, \A, and \B). 
The spectroscopic model created is used to evaluate CN as a molecular probe to constrain the variation of the proton-to-electron mass ratio; no overly promising sensitive transitions for extragalactic study were identified. 
\end{abstract}


\section{Introduction}

CN is a molecule ubiquitous in astrochemistry that has been well studied experimentally and theoretically. CN shows initial promise as a potential molecular probe to constrain the variation of the proton-to-electron mass ratio  \citep{19SyMoCu.CN, 20SyMc.RN}, however an energy spectroscopic model of CN - not currently available - is required to identify key transitions of enhanced sensitivity.

The cyano (CN) radical was the second observed molecular species in the interstellar medium \citep{40Mc.CN} and  observed extra-galactically by \cite{88HeMaSc.CN}. 
CN is one of the most widely distributed astrophysical molecules observed across the electromagnetic spectrum. For example, in the optical and visible, CN has been seen in  interstellar clouds \citep{40Mc.CN, ritchey2011interstellar}, carbon stars \citep{96BaStKe.CN}, and the Hale Bopp comet \citep{97WaSc.CN}. In infrared, CN has been seen in comets \citep{1881Hu.CN, 17ShKaKo.CN}, active galactic nuclei \citep{07RiPaRo.CN}, and sunspots \citep{73Ha.CN}. In the microwave region CN has been observed in the Orion nebula \citep{70JePeWi.CN} and throughout diffuse clouds \citep{78AlKn.CN}. CN provides insight into diverse astrophysical phenomena. CN stars (a peculiar type of carbon stars) have unusually strong CN peaks present in their spectra \citep{96BaStKe.CN}. CN is often used as a tracer of gas layers which are affected by photochemistry, and is predominantly found in regions exposed to ionizing stellar UV radiation \citep{07RiWaCo.CN}. The rotational temperature of CN is used to estimate the brightness of the cosmic microwave background \citep{11RiFeLa.CN}. 
CN is an important part of the HCN cycle, a key precursor in the synthesis of prebiotic molecules, such as nucleotides, amino acids, and lipid building blocks \citep{17FeKuKn.CN},. %

Modelling observations of astronomical or other gaseous environments accurately, and thus understanding these environments, requires high accuracy line lists - i.e. details of all the energy levels in a molecule and the strength of transitions between these levels. These in turn rely on high-quality experimental data. In the case of CN, we need to consider at least the three lowest lying electronic states, the \X{} ground state, the \A{} state at 9243 \cm{}, and the \B{} state at 25752 \cm{}, as astrophysical observations of interactions between all of these electronic bands have been observed \citep{19HaKaKo.CN}. All available experimental data was recently compiled for $^{12}$C$^{14}$N \citep{20SyMc.CN}, which then used the \Marvel{} \citep{marvel} procedure to extract empirical energy levels.

For CN, the most accurate available line list is the \Mollist{} data  \citep{14BrRaWe.CN}, which considers transitions between the three lowest electronic states of CN - i.e. the \X{}, \A{} and the \B{} states. Uses of the \Mollist{} line list have included; constraining the chemical evolution of the local disc with C and N abundances \citep{20BoDeMe.CN}, modelling the impact of chemical hazes in exoplanetary atmospheres \citep{21LaAr.CN}, probing interstellar clouds \citep{20WeSoSn.CN}, chemical abundance in stars that may harbour rocky planets from TESS data \citep{20TaMiDr.CN}, as well as observing the first \A-\X{} (0,0) band in the interstellar medium \cite{19HaKaKo.CN}. 

The \Mollist{} line list is computed using the so-called traditional model (or perturbative method), fitting experimental transition frequencies to a model Hamiltonian using PGopher \citep{western2017pgopher} to obtain a set of spectroscopic constants which are then used to predict unobserved line frequencies along with \abinitio{} dipole moments, see \cite{14BrRaWe.CN} for further details. The \Mollist{} traditional model interpolates very accurately but does not extrapolate well because it is based on perturbation theory \citep{20Be.CN}. The CN \Mollist{} line list has no published underlying energy spectroscopy model (i.e. set of potential energy  and coupling curves) suitable for testing the sensitivity of its molecular transitions to variation in the proton-to-electron mass ratio. 

Variational line lists, such as those developed by ExoMol \citep{20TeYuAl.CN} and TheoReTS \citep{16ReNiBa.CN}, are based on a spectroscopically fitted energy spectroscopic model - i.e. potential energy and coupling curves. One popular program to create a variational line list for diatomics is the nuclear motion program \Duo{} which variationally solves the nuclear motion Schrodinger equation for coupled electronic states \citep{16YuLoTe.mp2me, 17TeYu.exomol}. \Duo{} has been successfully utilised to generate spectroscopic data for over 15 diatomic molecules \citep{20TeYuAl.CN}. 
 
 These variational line lists "extrapolate more reliably because [they are] a more realistic and less empirical model" \citep{20Be.CN}, but are not as accurate for each individual vibronic band due to the reduced number of free parameters and increased physical constraints. No variational line list and thus no energy spectroscopic model exists for CN prior to this paper. This missing spectroscopic model meant that CN could not be rigorously evaluated as a potential molecular probe for proton-to-electron mass variation in \citet{19SyMoCu.CN}. 

Experimental accuracy for individual lines (often the strongest lines) can be achieved for traditional and variational line lists by replacing energy levels with experimentally-derived energy levels best obtained from a \Marvel{} inversion of all experimental transitions.  

Both traditional and variational line lists have their advantages. To take best advantage of all diatomic spectroscopic data (reviewed by \cite{21Mc.CN}), we propose here to use a novel trihybrid approach combining data from the traditional and variational approaches with the best available experimental data.  

This paper is organised as follows. In Section 2, we describe the creation of the energy spectroscopic model and analyse the resulting variationally predicted energy levels in comparison to \Mollist{} and \Marvel{} energy levels. 

In Section 3, we add the intensity spectroscopic model and combine previous line list approaches to develop and explore a novel \LLname{} line list. %
Finally, in Section 4, we use the newly generate energy spectroscopic model from Section 2 to consider CN as a potential probe to constrain the variation of the proton-to-electron mass ratio, and calculate the sensitivity coefficient of transitions generated by the final spectroscopic model.

\section{Energy Spectroscopic Model (ESM)}

\subsection{Construction}
An energy spectroscopic model (ESM) consists of potential energy curves (PECs) for each electronic state and coupling curves between those states. ESMs for diatomic molecules can be constructed using \Duo{} \citep{16YuLoTe.mp2me, 17TeYu.exomol}, where each PEC and coupling curves are represented by a mathematical functions. 

The parameters of these functions are fit using \Duo{} to minimise the difference between the Duo-predicted variational energy levels and the available \Marvel{} experimentally-derived energy levels \citep{20SyMc.CN}. \Duo{} uses a grid-based sinc DVR method to solve the coupled Schrodinger equation; for our calculations, we use grid of uniformly distributed 1001 points from 0.6-4.0 \AA. This fit is highly non-linear and so fitting is normally done iteratively, often considering just one electronic or vibronic state at a time. The $v$ = 14, 16, and 18 states of the \B{} state were  unweighted in the \Duo{} fit due to large perturbations. For further details of the fitting process, see \cite{16YuLoTe.mp2me, 17TeYu.exomol, Tennyson2016}. Of particular note for CN was the high vibrational and rotational quantum numbers for which experimental data is available; essentially removing the need for \abinitio{} data as a starting point. 

During the fitting process we identified a few misassignments in the \Marvel{} data. The transitions that these data was taken from were identified and removed; the updated \Marvel{} files are in the supplementary material.

\begin{table}
 \label{tab:PEC}
     \sisetup{round-mode=figures,scientific-notation = fixed, fixed-exponent = 0}
\caption{Fitted parameters of potential energy curves for CN \Duo{} spectroscopic model using extended Morse oscillator defined in \citet{duomanual}.  VE and AE are given in \cm{}, while the equilibrium bond lengths, RE, are given in \AA{}; all other parameters are dimensionless. PL = 4, PR = 4, NL = 4 for all states while NR = 8, 5 and 10 for the \X{}, \A{} and \B{} states respectively.}
\begin{tabular}{@{}lS[round-precision=6]S[round-precision=6]S[round-precision=6]@{}}
\toprule
 & \mc{1}{c}{\X{}}  & \mc{1}{c}{\A{}} & \mc{1}{c}{\B{}} \\ \midrule
VE & 0.000000E+00 & 9.246871E+03 & 2.575559E+04 \\
RE & 1.172719E+00 & 1.231350E+00 & 1.149790E+00 \\
AE & 6.361940E+04 & 6.361940E+04 & 8.284311E+04 \\
B0 & 2.539038E+00 & 2.406741E+00 & 2.793785E+00 \\
B1 & 1.983928E-01 & 1.222694E-01 & 3.948726E-01 \\
B2 & 1.908719E-01 & 1.551045E-01 & 5.930972E-02 \\
B3 & 2.411238E-01 & 8.099625E-02 & -1.812743E+00 \\
B4 & 3.699915E-01 & -1.104037E-01 & -1.941977E+00 \\
B5 & -1.365487E+00 & 7.013091E-01 & -3.247205E+00 \\
B6 & 2.545451E+00 &  & 8.027914E+00 \\
B7 & 0.000000E+00 &  & 0.000000E+00 \\
B8 & -8.652035E-02 &  & 2.485824E+00 \\
B9 &  &  & 0.000000E+00 \\
B10 &  &  & -5.030818E+00 \\ \bottomrule
\end{tabular}

\end{table}

\begin{figure}
    \centering
    \includegraphics[width = 0.48\textwidth]{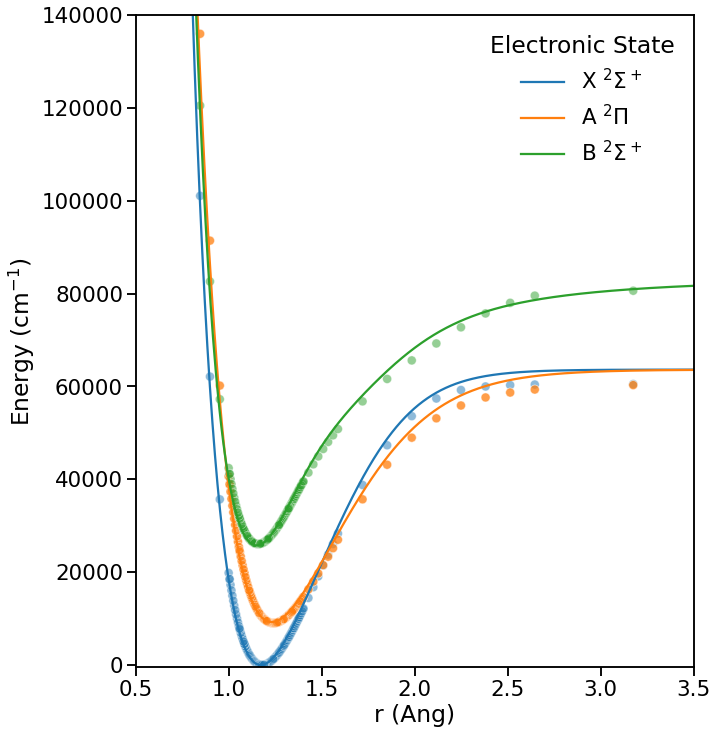}
    \caption{Potential energy curves of the 3 lowest electronic states of CN. The solid line shows the final curves fit using MARVEL energy levels, and the dots show the \abinitio{} calculations. }
    \label{fig:poten}
\end{figure}

\paragraph*{Potential Energy Curves:}

The potential energy curves (PEC) for the \X{}, \A{}, and \B{} electronic states were described using the Extended Morse Oscillator (EMO) \citep{99LeSeHi.CN} with standard parameterisation, detailed in \Cref{tab:PEC}. %
The \X{} and \A{} state have a common dissociation asymptote fixed to the experimental value of 63,619.4 \cm{} \citep{79HeHu.diatomic}, whereas the \B{} state has a higher dissociation limit fixed at 82,843.11 \cm{} \citep{18YiShSu.CN}. To ensure a sufficient fit, a significant amount of parameters were used to fit these states due to the large amount of experimental data including very high $v$ and high $J$ energy levels from \Marvel.   %

The resulting PECs are shown in \Cref{fig:poten}. Calculated \abinitio{} curves are provided for reference but not used in the construction of the ESM due to the wealth of available experimental energy levels. The fitted and \abinitio{} curves are very similar near the bottom of the potential well, with increasing deviation at larger $r$ because the \abinitio{} predicted dissociation energy differed from the experimental value. Our calculations used MRCI/aug-cc-pVTZ level with an (6,2,2,0) active space based on state-averaged (\X{}, \A{} and \B{} only) Complete Active Space Self-Consistent Field (SA-CASSCF) calculations. We note that similar \abinitio{} results are presented in, for example, \cite{11ShLiSu.CN, 18YiShSu.CN}; however, underlying raw data could not be obtained, highlighting the importance of data accessibility \citep{21Mc.CN}.

\begin{table}
     \sisetup{round-mode=figures,scientific-notation = fixed, fixed-exponent = 0}
\centering
\caption{Fitted constants, as defined in \citet{16YuLoTe.mp2me}, affecting energy levels of individual electronic states: diagonal spin– spin ($\lambda_\textrm{SS}$), spin–rotational ($\gamma_\textrm{SR}$), rotational Born- Oppenheimer breakdown term ($B_\textrm{rot}$), spin–orbit (SO), and lambda doubling ($\lambda_\textrm{p2q}$, $\lambda_\textrm{opq}$) constants in \cm. Off-diagonal spin–orbit and electronic angular momentum coupling terms respectively}
\label{tab:coup}
\resizebox{0.5\textwidth}{!}{
\begin{tabular}{lS[round-precision=3]S[round-precision=3]S[round-precision=3]}
\toprule
Diagonal  & {\X{}} & {\A{}}& {\B{}} \\
\cmidrule(r){2-4} 
SO, B0 & & -2.65E+01 \\
SO, B1 & & -7.22E+00 \\
$\lambda_\textrm{SS}$ & 0.100 & 0.100 & 1.0\\
$\gamma_\textrm{SR}$ & 1.31E-02 & -3.70E-03 & 1.83E-02 \\
$B_\textrm{rot}$ & 1.62E-03 &  -2.66E-03  & -3.23E-03 \\
$\lambda_\textrm{p2q}$ & &2.31E-02 &  \\
$\lambda_\textrm{opq}$  & & -2.77E-02 &  \\

\\

Off Diagonal & {X, A}  & {B, A} \\
\cmidrule(r){2-4} 

SO       & 1.88E+01  & 1.90E+00  \\
L+        & -3.68E-01 & -3.68E-01 \\
\bottomrule
\end{tabular}}

\end{table}

\paragraph*{Coupling:} 
The interaction of electronic states is described through a variety of coupling curves, all described using the Surkus-polynomial expansion \citep{16YuLoTe.mp2me} to ensure correct asymptotic behaviour. Various diagonal and off-diagonal terms were included as part of the fit to the \Marvel{} energy levels, described in \Cref{tab:coup}, with the most important term the diagonal spin-orbit coupling term for the \A{} state.

\begin{figure*}
    \centering
    \includegraphics[width = 0.90\textwidth]{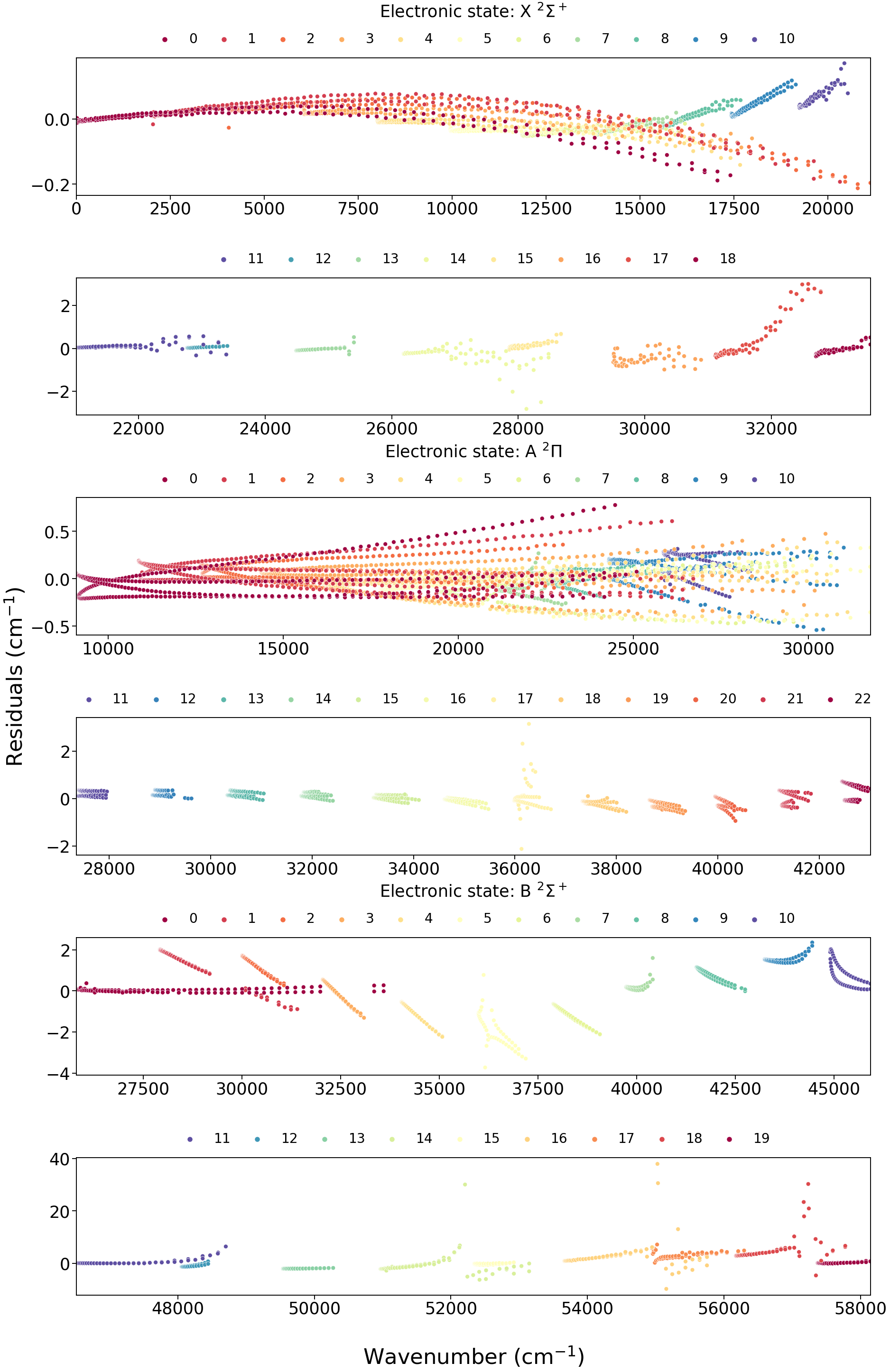}
    \caption{Residuals energy differences (in \cm) between the \Duo{} energies and the \MARVEL{} empirical energy levels, split into two subfigures for each electronic state. Note the differing vertical scales. }
    \label{fig:res_states_v}
\end{figure*}

\begin{table}
    \sisetup{round-mode=places,retain-explicit-plus}
\centering
\caption{Vibronic-scale breakdown of the average deviation of the \Duo{} and \Mollist{} energy levels  from the \Marvel{} empirical energy levels. All deviations are in \cm{}. }
\label{tab:RMSD}
\resizebox{0.48\textwidth}{!}{
\begin{tabular}{HrrS[round-precision=2]S[round-precision=2]S[round-precision=2]S[round-precision=2]S[round-precision=2]S[round-precision=2]}
\toprule
 & \multicolumn{1}{c}{} & \multicolumn{1}{c}{} & \multicolumn{3}{c}{\Duo{} (this work) - \Marvel{}} & \multicolumn{3}{c}{\Mollist{}   - \Marvel{}} \\ \midrule
 & \multicolumn{1}{c}{$v$} & \multicolumn{1}{c}{Max $J$} & \multicolumn{1}{c}{Mean} & \multicolumn{1}{c}{RMSD} & \multicolumn{1}{c}{|Max|} & \multicolumn{1}{c}{Mean} & \multicolumn{1}{c}{RMSD} & \multicolumn{1}{c}{|Max|} \\ 
 \cmidrule(r){4-6} \cmidrule(r){7-9}
& \mc{2}{l}{\textbf{\X{}}} \\
 & 0 & 97.5 & -0.009495 & 0.048411 & 0.1888 & -0.014865 & 0.017354 & 0.050417 \\
 & 1 & 99.5 & 0.020253 & 0.053666 & 0.1928 & -0.014414 & 0.01682 & 0.047269 \\
 & 2 & 97.5 & 0.00805 & 0.063984 & 0.2126 & -0.014684 & 0.018409 & 0.067733 \\
 & 3 & 81.5 & 0.003786 & 0.030592 & 0.097 & -0.013494 & 0.015391 & 0.034892 \\
 & 4 & 72.5 & -0.022725 & 0.035192 & 0.141 & -0.012947 & 0.015071 & 0.045676 \\
 & 5 & 60.5 & -0.035955 & 0.037488 & 0.0729 & -0.011156 & 0.012504 & 0.028125 \\
 & 6 & 48.5 & -0.03841 & 0.039854 & 0.0526 & -0.004736 & 0.009229 & 0.047625 \\
 & 7 & 36.5 & -0.028641 & 0.032933 & 0.048 & -0.007736 & 0.009091 & 0.031186 \\
 & 8 & 34.5 & 0.005169 & 0.026081 & 0.0589 & -0.008577 & 0.01027 & 0.034995 \\
 & 9 & 30.5 & 0.044384 & 0.054936 & 0.118 & -0.009899 & 0.010859 & 0.019801 \\
 & 10 & 27.5 & 0.067496 & 0.074281 & 0.1708 & -0.012525 & 0.016873 & 0.068955 \\
 & 11 & 36.5 & 0.098832 & 0.185389 & 0.5689 & 0.592017 & 1.887538 & 5.933572 \\
 & 12 & 19.5 & 0.049649 & 0.056052 & 0.1134 & -0.004243 & 0.015815 & 0.050708 \\
 & 13 & 23.5 & -0.037104 & 0.116005 & 0.5271 & -0.000012 & 0.099609 & 0.518402 \\
 & 14 & 37.5 & -0.375269 & 0.633757 & 2.8284 & 0.025254 & 0.173347 & 0.645563 \\
 & 15 & 22.5 & 0.155627 & 0.228814 & 0.6775 & 0.103539 & 0.154551 & 0.543499 \\
 & 16 & 29.5 & -0.508863 & 0.584798 & 0.9888 &  &  &  \\
 & 17 & 32.5 & 0.545994 & 1.223023 & 3.0129 &  &  &  \\
 & 18 & 23.5 & -0.07864 & 0.219089 & 0.5124 &  &  &  \\
 \vspace{-0.5em}\\
&  \mc{2}{l}{\textbf{\A{}}}&\\
 & 0 & 98.5 & -0.02449 & 0.197597 & 0.7764 & -0.014912 & 0.017492 & 0.055477 \\
 & 1 & 98.5 & 0.062223 & 0.179581 & 0.6066 & -0.013122 & 0.014935 & 0.027548 \\
 & 2 & 80.5 & 0.067546 & 0.135888 & 0.3589 & -0.012491 & 0.014719 & 0.040838 \\
 & 3 & 99.5 & 0.003343 & 0.148708 & 0.4733 & -0.018747 & 0.10703 & 1.13708 \\
 & 4 & 97.5 & -0.047244 & 0.155858 & 0.4127 & -0.004375 & 0.08442 & 0.980367 \\
 & 5 & 94.5 & -0.059818 & 0.166894 & 0.4539 & -0.038124 & 0.301993 & 3.142193 \\
 & 6 & 82.5 & -0.061279 & 0.170878 & 0.471 & -0.079183 & 0.409199 & 3.829591 \\
 & 7 & 37.5 & -0.045284 & 0.109901 & 0.2725 & -0.30002 & 1.152204 & 4.819375 \\
 & 8 & 41.5 & 0.018606 & 0.088537 & 0.2886 & 0.269988 & 1.047497 & 7.608423 \\
 & 9 & 65.5 & 0.062874 & 0.188273 & 0.541 & -0.026939 & 0.060429 & 0.333736 \\
 & 10 & 39.5 & 0.140487 & 0.187436 & 0.3142 & -0.005066 & 0.027613 & 0.279478 \\
 & 11 & 19.5 & 0.211977 & 0.238453 & 0.3508 & -0.005493 & 0.00774 & 0.017516 \\
 & 12 & 22.5 & 0.22102 & 0.249403 & 0.3689 & -0.004789 & 0.009372 & 0.025182 \\
 & 13 & 21.5 & 0.195882 & 0.226677 & 0.3569 & -0.005603 & 0.008765 & 0.02655 \\
 & 14 & 20.5 & 0.143333 & 0.176733 & 0.2923 & -0.006049 & 0.00864 & 0.021671 \\
 & 15 & 23.5 & 0.050293 & 0.106314 & 0.1935 & -0.003328 & 0.022033 & 0.188036 \\
 & 16 & 24.5 & -0.083048 & 0.130675 & 0.4432 & -0.007657 & 0.010716 & 0.042111 \\
 & 17 & 22.5 & 0.083696 & 0.670126 & 3.1544 & 0.011811 & 0.156897 & 0.601659 \\
 & 18 & 23.5 & -0.220773 & 0.251927 & 0.5613 & -0.001997 & 0.042054 & 0.266725 \\
 & 19 & 22.5 & -0.297024 & 0.329162 & 0.6222 & -0.009333 & 0.013415 & 0.058162 \\
 & 20 & 19.5 & -0.256248 & 0.344587 & 0.9432 & -0.009121 & 0.012318 & 0.036189 \\
 & 21 & 21.5 & 0.029403 & 0.270282 & 0.4048 & -0.009728 & 0.022321 & 0.125458 \\
 & 22 & 20.5 & 0.298623 & 0.450499 & 0.7231 & -0.001442 & 0.01138 & 0.042571 \\
 \vspace{-0.5em}\\
&  \mc{2}{l}{\textbf{\B{}}}&\\
 & 0 & 63.5 & 0.024335 & 0.088594 & 0.3713 & -0.014476 & 0.029433 & 0.155259 \\
 & 1 & 41.5 & 1.148954 & 1.46008 & 2.0129 & -0.019459 & 0.024899 & 0.065918 \\
 & 2 & 23.5 & 1.200113 & 1.277858 & 1.718 & -0.020308 & 0.02686 & 0.065065 \\
 & 3 & 23.5 & -0.103533 & 0.57159 & 1.3074 & 0.106561 & 0.877309 & 6.011292 \\
 & 4 & 23.5 & -1.155134 & 1.270891 & 2.2323 & -0.013907 & 0.017788 & 0.045733 \\
 & 5 & 24.5 & -1.911794 & 2.092138 & 3.7222 & 0.015294 & 0.367964 & 1.71131 \\
 & 6 & 25.5 & -1.189353 & 1.277696 & 2.1203 & -0.007871 & 0.00932 & 0.022564 \\
 & 7 & 19.5 & 0.223138 & 0.346195 & 1.5993 & -0.046895 & 0.118713 & 0.519999 \\
 & 8 & 26.5 & 0.724938 & 0.801644 & 1.1554 & -0.011399 & 0.01306 & 0.036121 \\
 & 9 & 26.5 & 1.550311 & 1.563056 & 2.3519 & -0.026736 & 0.074577 & 0.423766 \\
 & 10 & 24.5 & 0.836188 & 1.033244 & 2.0306 & -0.004191 & 0.088118 & 0.333109 \\
 & 11 & 36.5 & 0.507172 & 1.232155 & 6.4299 & -0.05214 & 0.132541 & 0.628043 \\
 & 12 & 15.5 & -0.894435 & 1.014615 & 1.2766 & -0.044951 & 0.11093 & 0.480357 \\
 & 13 & 21.5 & -1.995677 & 1.998493 & 2.1236 & -0.004988 & 0.017847 & 0.046609 \\
 & 14 & 37.5 & -0.739462 & 4.353172 & 30.0771 & -0.082113 & 0.685225 & 3.159693 \\
 & 15 & 19.5 & -0.197571 & 0.307877 & 0.6223 & 0.075576 & 0.134084 & 0.503881 \\
 & 16 & 37.5 & 2.662596 & 6.454269 & 37.9468 &  &  &  \\
 & 17 & 30.5 & 3.344022 & 3.611799 & 7.1818 &  &  &  \\
 & 18 & 33.5 & 5.201178 & 7.302556 & 30.2827 &  &  &  \\
 & 19 & 23.5 & 0.223726 & 0.319211 & 0.9396 &  &  &  \\ \bottomrule
\end{tabular}}
\end{table}

\subsection{Analysis}

\paragraph*{Comparison of variational \Duo{} with experimentally-derived \Marvel{} energy levels:}

\Cref{fig:res_states_v} visually compares the variational \Duo{} energy levels to the experimentally-derived \Marvel{} energy levels as a function of energy. The mean, RMSD, and maximum absolute deviation for each vibronic level collated in Table \ref{tab:RMSD}. Overall, the \X{} and \A{} states are very well fitted with an overal RMSD of 0.29 \cm{} and 0.21 \cm{} respectively.  The \B{} state is more problematic, but the rmsd is still 2.97 \cm{}, with errors dominated by the weaker fits in the strongly perturbed higher vibronic levels. %

\begin{table*}
    \centering
    \caption{Spectroscopic equilibrium parameters for the 3 electronic states in the \Duo{} spectroscopic model, compared with select values from the literature. }%
    \label{tab:eqparm}
\begin{tabular}{@{}llrrrrr@{}}
\toprule
Electronic State & Parameter & This work  & \cite{14BrRaWe.CN}$^*$ & \cite{18YiShSu.CN} & \cite{79HeHu.diatomic} & \cite{09BaRiPe.CN} \\ \midrule
\X{}             & $T_e$        & 0.0000     & 0                  & 0                  & 0                      & 0                  \\
                 & $\omega_e$        & 2068.7230  & 2068.68325         & 2069.26            & 2068.59                & 2068.65            \\
                 & $\omega_e\chi_e$      & 13.1808    & 13.12156           & 10.231             & 13.087                 & 13.097             \\
                 & $B_0$        & 1.8968     & 1.89978            & 1.9013             & 1.8997                 & 1.899783           \\
                 & $R_e$        & 1.1727     & 1.1718             & 1.1714             & 1.1718                 & -                  \\
                 & $\alpha_e$        & 0.0174     & 0.01738           & 0.01719            & 0.01736                & 0.01737            \\
\A{}             & $T_e$        & 9247.2180  & 9243.2959          & 9109.95            & 9245.28                & 9240               \\
                 & $\omega_e$        & 1812.8320  & 1813.288           & 1814.75            & 1812.56                & 1813.26            \\
                 & $\omega_e\chi_e$      & 12.5944    & 12.77789           & 13.053             & 12.6                   & 12.7687            \\
                 & $B_0$        & 1.7205     & 1.7158             & 1.7174             & 1.7151                 & 1.7159             \\
                 & $R_e$        & 1.2314     & 1.233044           & 1.2324             & 1.2333                 & -                  \\
                 & $\alpha_e$        & 0.0172     & 0.01725           & 0.01708            & 0.01708                & 0.017167           \\
\B{}             & $T_e$        & 25755.5900 & 25752.59           & 25776.42           & 25752                  & 25752              \\
                 & $\omega_e$        & 2156.2630  & 2162.223           & 2163.04            & 2163.9                 & 2161.46            \\
                 & $\omega_e\chi_e$      & 16.2659    & 19.006             & 14.789             & 20.2                   & 18.219             \\
                 & $B_0$        & 1.9732     & 1.96797            & 1.9554             & 1.973                  & 1.96891            \\
                 & $R_e$        & 1.1498     & 1.15133            & 1.151              & 1.15                   & -                  \\
                 & $\alpha_e$        & 0.0178     & 0.0188            & 0.01908            & 0.023                  & 0.020377           \\ \bottomrule
\end{tabular}
$^*$ $\alpha_e$ for \citet{14BrRaWe.CN} was the opposite sign for all three states than other sources; we assume a definition difference. 
\end{table*}

\begin{figure}
    \centering
    \includegraphics[width = 0.48\textwidth]{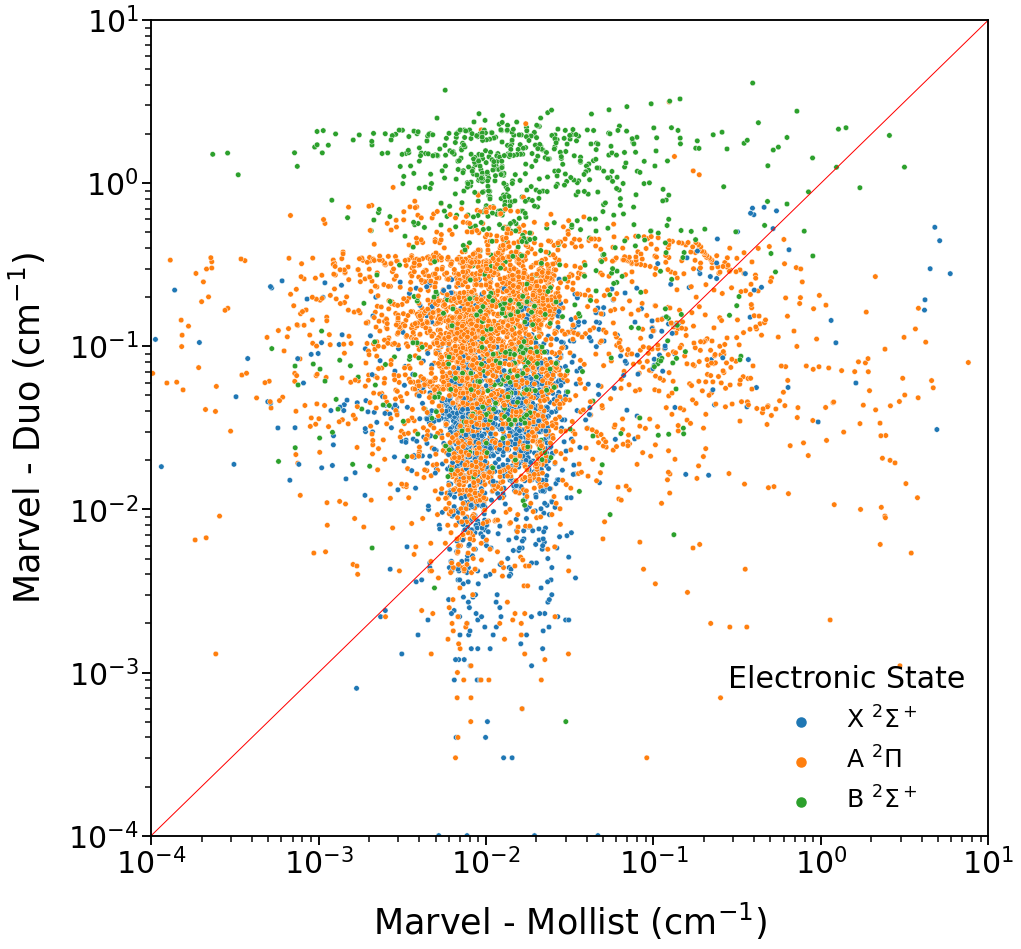}
    \caption{Comparison of the residuals with \Marvel{} empirical energy levels from \Mollist{} and the \Duo{} ESM.}
    \label{fig:energy_mol_duo}
\end{figure}

The first 2 panels for figure \ref{fig:res_states_v} show the deviation of the \X{} state. We can see that the residuals of the \X{} state are fairly consistent across each vibrational level, with the exception of $v$ = 17 which has RMSD greater than 1 \cm{}. The maximum absolute deviation from the \Marvel{} energies in the \X{} state is 3.01 \cm{} in the $v$ = 17 state. 

The \A{} state, shown in panels 3 and 4 show  consistent errors for all vibronic states, with no vibronic level having an RMSD greater than 1 \cm{}, even the $v$ = 17 state, which visibly looks scattered only has a rsmd of 0.67 \cm. The maximum deviation occurs in the $v$ = 17 state, with an absolute deviation of 3.15 \cm. We see the diverging deviation in the lower vibronic levels of the \A{} state, while small, suggest an issue with the spin-spin coupling.

In the bottom two panels we see the deviation in the \B{} state. While the \B{} state is much higher in energy the deviation from the \Marvel{} energy levels seems especially large, and scattered. Over half of the vibronic levels in the \B{} state have a RMSD greater than 1 \cm{}, and all bar four vibonic levels have an RMSD greater than 0.5 \cm{}. The $v$ = 16 to $v$ = 18 have RMSD greater than 2.0 \cm. The maximum deviation of 37.95 \cm{} in the \B{} state occurs at $J$ = 30.5 and $v$ = 16. The poor performance for high $v$ levels in the \B{} state is probably caused by perturbations that are not represented in our spectroscopic model, for example with the spectroscopically dark \asig{} state. %

\paragraph*{Comparison of variational and perturbative energy levels:} 
We compare the \Mollist{} traditional/ perturbative energy levels with our variational energy levels by comparing the predicted energy levels against 6122 \Marvel{} empirical energy levels, visually in \Cref{fig:energy_mol_duo} and quantiatively in \Cref{tab:RMSD}. 86\% of the energy levels \Mollist{} are closer to the \MARVEL{} empirical energy levels than those produced using the ESM in \Duo{}. The improved behaviour arises because the \Mollist{} perturbative method uses individual descriptions of each vibronic level enabling easier treatment of perturbations than a physically self-consistent ESM that must describe all vibronic levels. \Cref{tab:RMSD} does show some vibronic bands with significant \Mollist{} errors, probably energies that were not included in the \Mollist{} initial fit.   %

\paragraph*{Spectroscopic constants:}

\Cref{tab:eqparm} compares the equilibrium spectroscopic parameters calculated by \Duo{} for our fitted potential energy curves against existing data \citep{18YiShSu.CN, 11ShLiSu.CN}, including those used to construct \Mollist. 
The parameters for the \X{} state agree very well across all sources, with a small difference in $\omega_e$ and $\omega_e \chi_e$ in the computational values of \cite{18YiShSu.CN}. The \A{} state has similar agreement across the sources, with a small difference of $T_e$ values, but a much closer consensus on the $\omega_e$ and $\omega_e \chi_e$ values. The \B{} state has the most disagreement across its equilibrium parameters. Our values for $\omega_e$ differ the most, and we see disagreement across all sources for $\omega_e \chi_e$. The difference between the \cite{14BrRaWe.CN} spectroscopic constants are our own are unlikely to be the primary reason for the differences in our predicted energy levels. Instead, the defining difference is likely to be the use of band constants in \Mollist{} line list.

\begin{figure*}
    \centering
    \includegraphics[width = 0.98\textwidth]{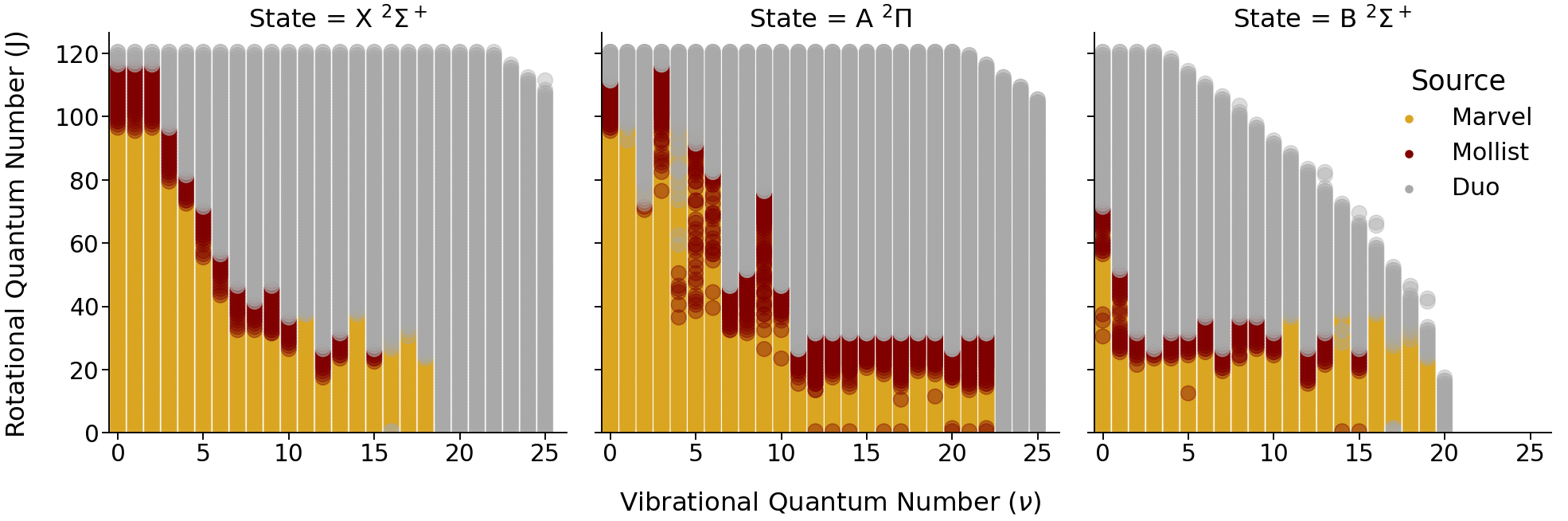}
    \caption{Distribution of sources used to create the final states file broken down across electronic state, rotational quantum number ($J$), and vibrational quantum number ($v$), capped at $v = 25$ for brevity. }
    \label{fig:statesbar}
\end{figure*}

\begin{table*}
\caption{Extract from the state file for \ce{^{12}C^{14}N}. Full tables are available at \url{www.exomol.com} and in the Supporting Information.}
    \label{tab:states}
\resizebox{0.98\textwidth}{!}{
\begin{tabular}{rrrrrrccccrrrrcrrr}
\toprule
$n$&   $\tilde{E}$ &  $g_{\rm tot}$ &    $J$ &  unc & $\tau$ &    $g$ & $+/-$ & $e/f$ & State &  $v$ &  $\Lambda$ &  $\Sigma$ &  $\Omega$ & Source &   $\tilde{E}_{\rm \Duo}$\\
\midrule
1&	0.000000&	6&	0.5	&0.001293&	-1.00E+00&	2.002305&	+&	e&	X&	0&	0&	0.5&	0.5&	M&	0.000000\\
102	&3.777245	&6	&0.5&	0.003132&	9.99E+04&	-0.667444&	-&	f&	X&	0&	0&	-0.5&	-0.5&	M&	3.775090\\
81&	52340.028680&	6&	0.5&	0.500000&	1.27E-07&	2.002272&	+&	e&	B&	15&	0&	0.5&	0.5	&P	&52340.262410\\
132	&28904.950010&	6&	0.5	&0.500000	&5.19E-06	&-0.000931	&-	&f&	A&	12&	-1	&0.5&	-0.5&	P	&28904.596730\\
78&	51326.871830&	6&	0.5	&1.000000&	1.51E-05&	-0.000658&	+&	e&	A&	30	&1&	-0.5&	0.5	&D	&51326.871830\\
79	&51797.529830&	6	&0.5&	1.000000&	9.39E-02&	2.002186&	+&	e&	X&	32&	0	&0.5&	0.5	&D	&51797.529830\\
\bottomrule
\end{tabular}
}
\begin{tabular}{cll}
\\
  Column       &    Notation      &      \\
\midrule
   1 &   $n$   &       Energy level reference number (row)    \\
   2 & $\tilde{E}$        &       Term value (in \cm) \\
   3 &  $g_{\rm tot}$     &       Total degeneracy   \\
   4 &  $J$    &       Rotational quantum number    \\
5 & unc & Uncertainty (in \cm) \\
  6 & $\tau$ & Lifetime (s) \\
   7 & $g$ & Land\'e factors \\
   8 & $+/-$ & Total parity  \\
   9 & $e/f$ & Rotationless parity \\
   10 & State & Electronic state \\
   11 & $v$ & State vibrational quantum number \\
  12 &  $\Lambda$ &   Projection of the electronic angular momentum \\
 13 & $\Sigma$ &   Projection of the electronic spin \\
14 & $\Omega$ & Projection of the total angular momentum ($\Omega=\Lambda+\Sigma$) \\
15 & Source & Source of term value; M = \Marvel, P = \Mollist{} \citep{14BrRaWe.CN}, D = \Duo{} \\
16 & $\tilde{E}_{\rm \Duo}$ & Energy from \Duo{} spectroscopic model \\
\bottomrule
\end{tabular}
\end{table*}

\section{Trihybrid line list}

A line list contains two files; a states (.states) file which lists all quantum states with their energies and quantum numbers and a transitions (.trans) file which details the strength of the transitions between states. %

\subsection{Energy Levels}
Here we present a novel trihybrid approach for the construction of the final states file that combines energy levels from different sources to produce the most accurate and complete line list possible with current data and techniques. 

The initial states file with 28,004 energy levels is produced with \Duo{} using the variational approach with the ESM from Section 2. For each quantum state, the \Duo{} energy is replaced when possible preferentially by \Marvel{} energies (6,864 levels) or otherwise by \Mollist{} energies (1,574 levels) where available. For the three electronic states considered in this line list, we show the distribution of sources used in our final states file in  \Cref{fig:statesbar}. Relatively few \Mollist{} energies are included, probably due to the breadth of energy levels empirically determined in our \Marvel{} study \citep{20SyMc.CN}. 

Extracts of our final states files along with column descriptors, are shown in  \Cref{tab:states}.

\begin{figure}
    \centering
    \includegraphics[width=0.49\textwidth]{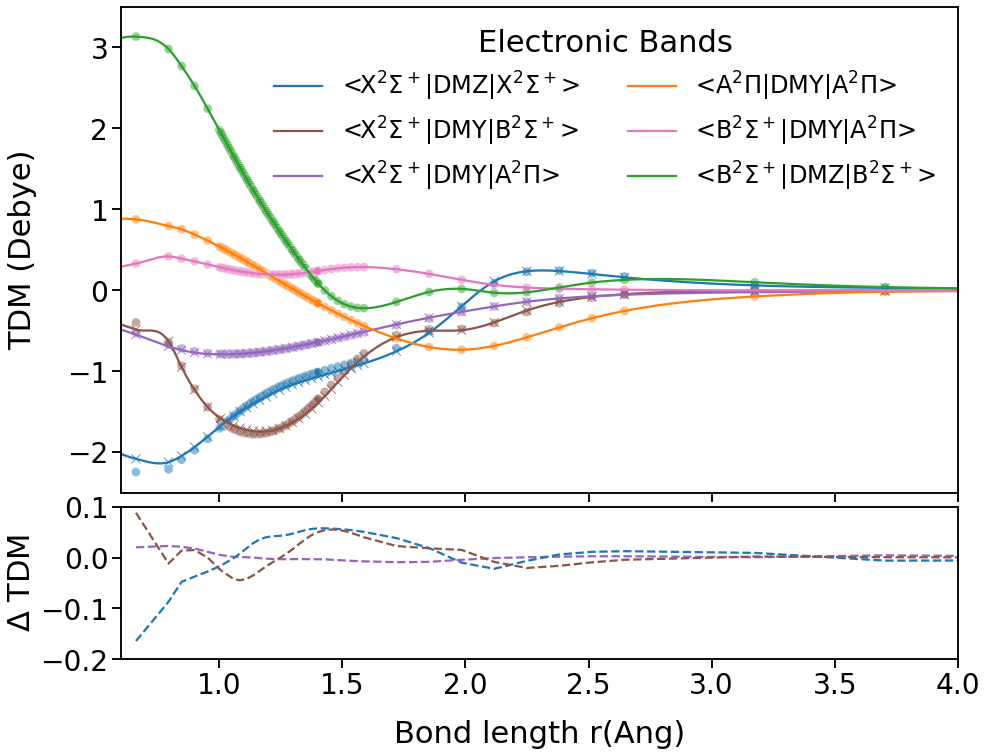}
    \caption{Dipole moment curves involving the three lowest electronic states of CN. The solid curves show the fitted \Duo{} curves, the circles are \abinitio{} calculations done in this work, and the crosses are the \abinitio{} data from \Mollist{} \citep{14BrRaWe.CN}}
    \label{fig:TDM}
\end{figure}

\subsection{Intensity Spectroscopic Model}
The intensity spectroscopic model consists of diagonal and off-diagonal (transition) dipole moments curves\alert{, calculated from high-level \abinitio{} methods then extrapolated within \Duo{}.}

\alert{Using our state-averaged MRCI/aug-cc-pVTZ calculation methodology discussed above, we calculated all relevant dipole moment curves, shown as circles in \Cref{fig:TDM}. This figure also shows the larger basis set results from \cite{14BrRaWe.CN} for the \X{}-\X{}, \A-\X{} and \B-\X{} curves as crosses. For our line list, we chose to use these larger basis set calculations for these curves, but note that there is only modest differences between the two \abinitio{} methods (bottom subfigure); the modest impact of this choice on our final line list is discussed in the Supporting Information. We input our selected dipole moment data points as a grid into \Duo{}, which fits a curve to these values. }  %

\alert{We validate our \abinitio{} results by comparing against experimental dipole moments. For the \X{} and \B{} states respectively, the experimental values are 1.45 $\pm$ 0.08 D and 1.15 $\pm$ 0.08 D \citep{68ThDa.CN}, similar to our equilibrium dipole moment of 1.34 D and 1.17 D.}%

\begin{table}
\caption{\label{tab:trans} Extract from the transition file for \ce{^{12}C^{14}N}. Full tables are available from \url{www.exomol.com} and in the Supporting Information.}
\centering
\begin{tabular}{rrrr}
\toprule
$f$  &  $i$  & $A({\rm f}\leftarrow {\rm i})$ / s$^{-1}$  \\
\midrule
181 &           1 &  0.000010 \\
182 &           1 &  3.080700 \\
183 &           1 &  0.216810 \\
184 &           1 &  0.002965 \\
185 &           1 &  0.004063  \\
\bottomrule
\end{tabular}

\noindent
 $f$: Upper (final) state counting number;
 
$i$: Lower (initial) state counting number;

$A({\rm f}\leftarrow {\rm i})$:  Einstein $A$ coefficient in s$^{-1}$.

\end{table}

\subsection{Transitions file}
The final line list contains 2,285,103 transitions, covering the wavenumber range 0-60,000 \cm{}. The transitions file is produced by combining the rovibronic wavefunctions produced by the energy spectroscopy model with the intensity spectroscopic model. Transitions were calculated for states with lower energies up to 30,000 \cm{} as this contains 99 \% of the total population at 5000 K. We have used a vibrational basis set up to $v$ = 120 for each electronic state, and calculated rotational levels up to $J$ = 120.5.

Extracts of our final transitions files along with column descriptors are shown in \Cref{tab:trans}. 

\begin{figure}
    \centering
    \includegraphics[width = 0.48\textwidth]{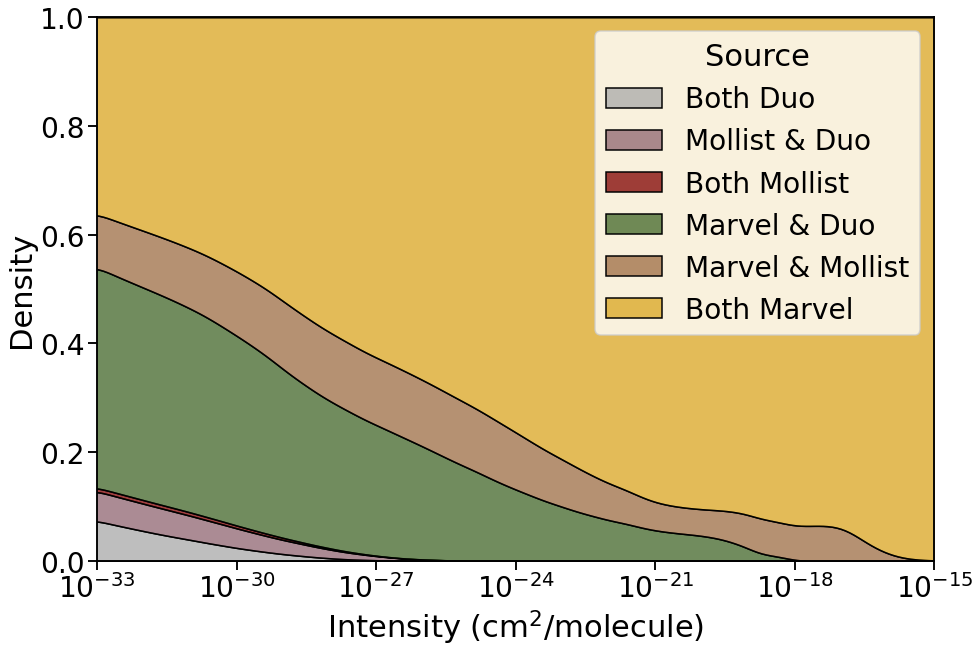}
    \caption{\alert{Cumulative} density distribution of sources of energy levels in transitions across intensity\alert{, i.e. at each vertical slice, all transitions with intensities at or above the intensity at that x axis point are considered.}  }
    \label{fig:transsources}
\end{figure}

90\% of strong transitions (defined as those with an intensity above $10^{-20}$ cm$^{2}$ molecule$^{-1}$ at 1000 K) have wavenumbers fully computed from \Marvel{} energy levels; they are thus highly reliable and the line list is very suitable for use to detect molecules through high-resolution cross-correlation techniques in exoplanets. The remaining 10\% of strong transitions contain one \Marvel{} energy level, with the other coming from the \Mollist{} energy levels half of the time, and \Duo{} predicted energy level the other half. Even considering all transitions with intensities over 10$^{-33}$ cm$^2$/molecule, fully \Marvel-ised transitions still make up 36\% of the distribution, with only 13\% of transition not containing any \Marvel{} energy level. \alert{Across all 2,285,103 transitions generated in the Trihybrid line list we see that 463,950 of them are fully \Marvel-ised, giving them pseudo-experimental accuracy. In comparison the original \Mollist{} line list includes 22,044 experimental transitions.}

\Cref{fig:transsources} provides a more in-depth analysis by visually quantifying the source of the Trihybrid transition wavenumbers as a function of the cumulative transition intensity at 1000 K. Fully \Marvel-ised  transitions dominate in all intensity windows above 10$^{-25}$ cm$^2$/molecule.

\subsection{Isotoplogues}
Full spectroscopic models and line list have been generated for three isotopologues of CN (\ce{^13C^14N}, \ce{^12C^15N}, and \ce{^13C^15N}), and the states and trans file have been included in the supplementary information. The states files for these isotopologues have been psudo-hybridised, as is standard for ExoMol isotopologue models \citep{marveliseisotopologues}, by shifting the energy levels of the isotopologues by the deviation between the main isotopologue 
\Duo{} and \Marvel{} or \Mollist{} energy, i.e. $E_\textrm{new}^\textrm{iso} = E_\textrm{Duo}^\textrm{iso} + (E_\textrm{trihybrid}^\textrm{main} - E_\textrm{Duo}^\textrm{main}) $ \Mollist{} computed line lists for \ce{^13C^14N}, \ce{^12C^15N} are also available from \cite{14SnLuRa.CN}.

\subsection{Results and Discussion}

\subsubsection{Partition function}
The partition function for CN was calculated using ExoCross \citep{18YuAlTe.exocross} across 10-7000 K in steps of 10 K using the following equation:

\begin{equation}
    Q(T) = \sum_n g_n^{tot} \exp^{-c_2 \Tilde{E_n} T},
    \label{equ:part}
\end{equation}

where $g_n^{tot}$ is total degeneracy, $g_n^{tot} = g_n^{ns}(2J_n + 1)$, and $g_n^{ns}$ is the `physics' interpretation of the nuclear spin statistical weight factor \citep{20PaYuTe.CN}, $c_2 = hc/kB$ is the second radiation constant (cm K), $\Tilde{E}_i = E_i/hc$ are the energy term values (\cm), taken from the states file, and $T$ is the temperature in K. 

\begin{figure}
    \centering
    \includegraphics[width = 0.49\textwidth]{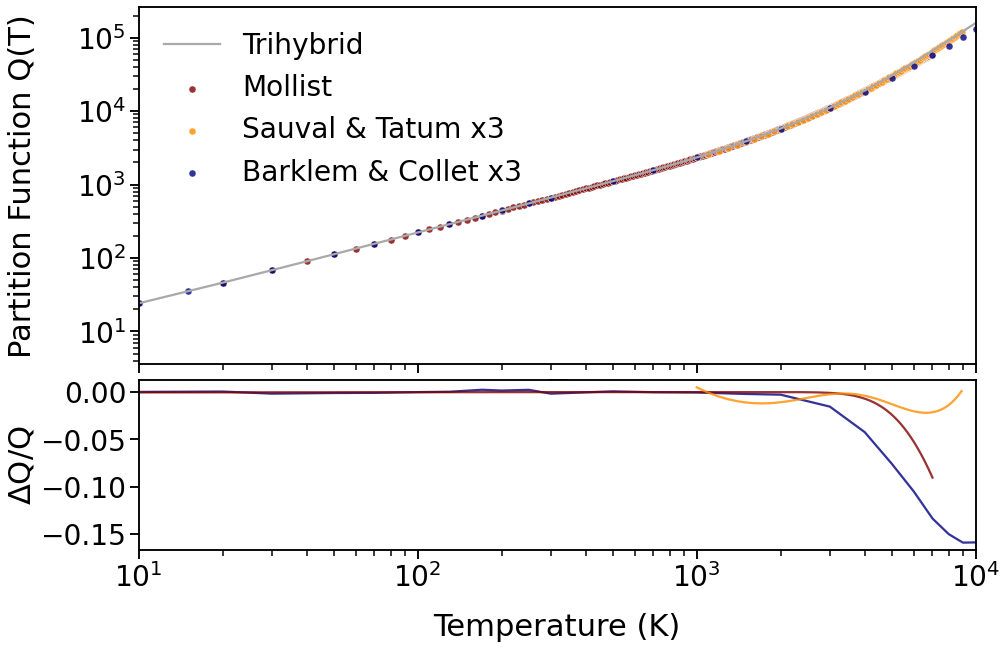}
    \caption{Comparison of the partition function for CN. Upper panel shows the partition function. Lower panel shows the relative deviation of each additional partition function to that from our hybrid states file. (Q(Other)-Q(Trihybrid))/Q(Trihybrid)}
    \label{fig:partfunc}
\end{figure}

We compare our partition function to that of Sauval and Tatum \citep{84SaTa.CN}, Barklem and Collet \citep{16BaCo.CN}, and the \Mollist{} line list, as shown in figure \ref{fig:partfunc}. To make a direct comparison between our partition functions and those from Sauval \& Tatum and Barklem \& Collet we have multiplied their values by the nuclear spin statistical weight for CN; $g_n^{ns} = 3$ to be in the 'physics` convention which is used by ExoMol. All four partition functions show a strong agreement with a maximum relative deviation from the Trihybrid partition function of 0.16 from Barklem \& Collet at 10,000 K. 

\begin{figure}
    \centering
    \includegraphics[width = 0.5\textwidth]{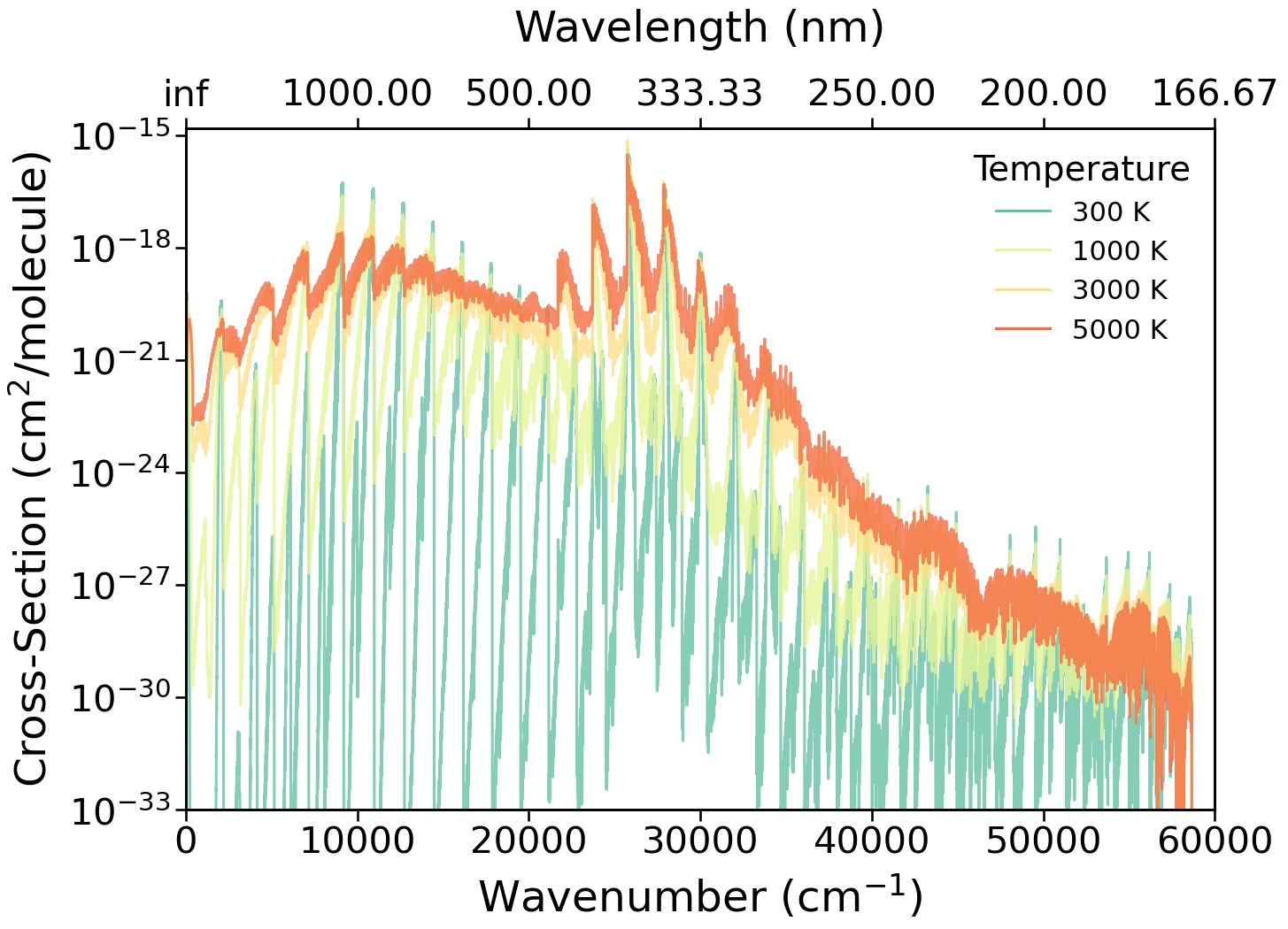}
    \caption{Cross section of CN up to 60,000 \cm{} for temperatures between of 300, 1000, 3000, and 5000 K with HWHM at 2 \cm. }
    \label{fig:XSec_temps}
\end{figure}

\begin{figure*}
    \centering
    \includegraphics[width = \textwidth]{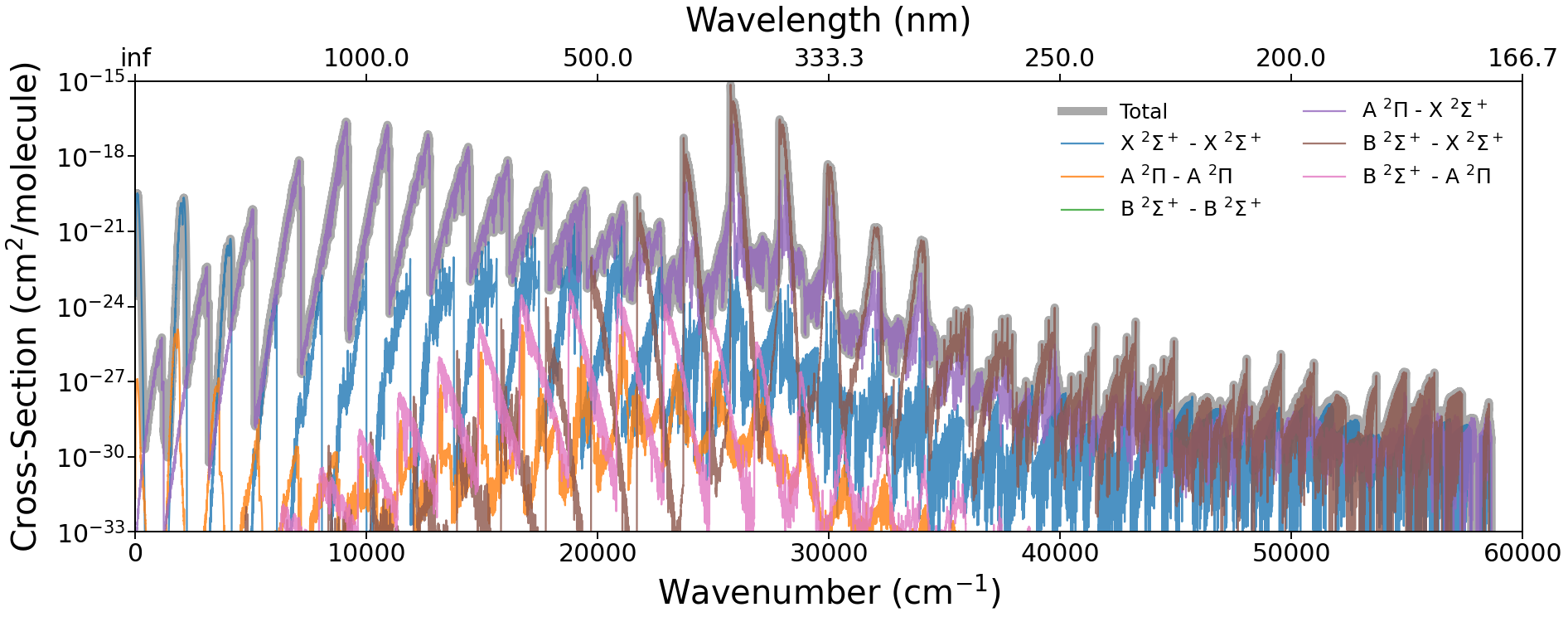}
    \caption{Cross section of CN from 0 - 60,000 \cm{}  at 1000 K with HWHM at 2 \cm, divided into component electronic bands. }%
    \label{fig:XSec_bands}
\end{figure*}

\begin{figure*}
    \centering
    \includegraphics[width = \textwidth]{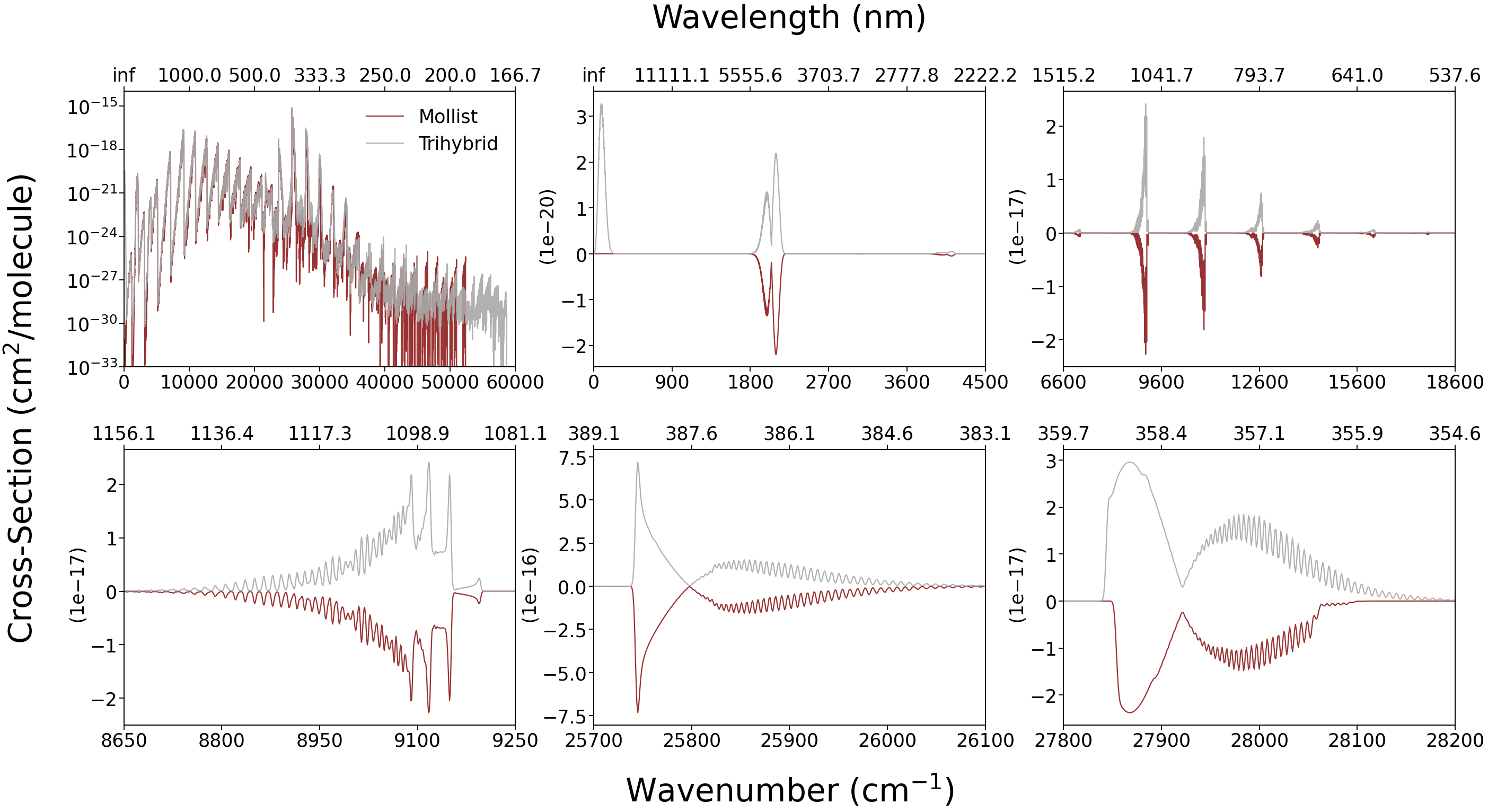}
    \caption{Comparison of the MARVELised \Mollist{}linelist cross section compared to the hybrid \Duo{} line list at 1000 K, at different wavenumber ranges.}
    \label{fig:Mol-hybrid-XSec}
\end{figure*}

\subsubsection{Cross sections}

\paragraph*{Overview}
We computed the absorption cross-section at temperatures of 300 K, 1000 K, 3000 K, and 5000 K, using a guassian line profile with a half width half maximum (HWHM) of 2 \cm{} in ExoCross, shown in figure \ref{fig:XSec_temps}. The spectra gets less defined with the increase in temperature, as expected. Above 35,000 \cm{} we see a very broadened cross section, losing almost all form at higher temperatures. 

Decomposing the absorption cross section at 1000 K into the bands in figure \ref{fig:XSec_bands} we see a strong dominance across all wavenumbers from the \A-\X{} band. The \X-\X{} band sees a strong occurrence in the microwave region before dropping off significantly, while the \B-\X{} bands is very diminished at lower frequencies, but taking dominance above 25,0000 \cm. The strongest feature of the spectra has a cross sectional intensity up to 10$^{-15}$ cm$^{2}$ molecule$^{-1}$ in the visible region, however there are strong ($<10^{-20}$ cm$^{2}$ molecule$^{-1}$) peaks within the infrared, microwave, and ultraviolet regions as well.

\paragraph*{Comparison with the \Mollist{} line list}
In the first subfigure of figure \ref{fig:Mol-hybrid-XSec} we compare the full cross section of the \MOLLIST{} line list and our hybrid line list from 0 - 60, 000 \cm{} modelled at 1000 K with a HWHM of 2 \cm{} using ExoCross. We see the completeness that is gained from the \Duo{} addition to the hybrid line list. The \Mollist{} line list has a similar cross section, especially at lower wavenumbers, however decomposes with an increase in wavenumber. We can see the \Mollist{} line list doesn't extend much further than 50,000 \cm, whereas the Trihybrid data extends much more smoothly out to 60,000 \cm. The other panels of figure \ref{fig:Mol-hybrid-XSec} compare the \Mollist{} cross sections and \LLname{} line list at a selection of key features. At wavenumber less than 4500 \cm{} we can see that the \Mollist{} line list does not contribute at all to the low frequency rotation lines, and while the \LLname{} does not include hyperfine transitions, it does still consider these rotational lines. The main infrared band is clearly matched well around 2000 \cm{}. The \A{}-\X{} bands shown between 6600 and 18600 \cm{} are also well matched. This is highlighted when we consider the main \A-\X{} (0,0) band around 9000 \cm{}, with peak positions and intensities matching extremely well. We begin to see some more obvious discrepancies between \Mollist{} and \LLname{} at higher energy. The main feature of the spectra (the \B{} - \X{} band) is matched pretty well at 25750 \cm{} but deviates above 27800 \cm{} due to the lower number of rotational energy levels included in the \Mollist{} line list.  \alert{Additional comparisons are provided in the Supporting Information. }

\begin{table*}
\centering
\caption{Comparison for lifetimes (ns) for vibrational states of the \A{} and \B{} electronic states. }
\label{tab:lifetimes}
\begin{tabular}{@{}lrrrrrr@{}}
\toprule
& \multicolumn{2}{c}{Line List} & \multicolumn{2}{c}{Experimental} & \multicolumn{2}{c}{Theory} \\ \midrule
v & This Work & \Mollist{} & \cite{92LuHuHa.CN} & \cite{78DuErLa.CN} & \cite{18YiShSu.CN} & \cite{84LaGaRo.CN} \\
\A{} &&&&&& \\
0 & 10735 & 11185 & - & - & 9980 & 11300 \\
1 & 9457 & 9680 & - & - & 9900 & 9600 \\
2 & 8515 & 8595 & 6960 $\pm$ 300 & 3830 $\pm$ 500 & 9670 & 8400 \\
3 & 7797 & 7785 & 5090 $\pm$ 200 & 4050 $\pm$ 400 & 8760 & 7600 \\
4 & 7234 & 7165 & 3830 $\pm$ 200 & 3980 $\pm$ 400 & 8040 & 6900 \\
5 & 6784 & - & 3380 $\pm$ 200 & 4200 $\pm$ 400 & 7400 & 6400 \\
6 & 6419 & - & 2260 $\pm$ 200 & 4350 $\pm$ 400 & 6960 & 6000 \\
7 & 6131 & - & 1840 $\pm$ 300 & 4350 $\pm$ 400 & 6540 & 5700 \\
8 & 6054 & - & - & 4500 $\pm$ 400 & 6190 & 5400 \\
9 & 5659 & - & - & 4280 $\pm$ 400 & 5890 & 5200 \\
10 & 5472 & - & - & 4100 $\pm$ 400 & 5640 & - \\
11 & 5318 & - & - & - & 5430 & - \\
12 & 5188 & - & - & - & 5270 & - \\
13 & 5077 & - & - & - & 5120 & - \\
 &  &  &  &  &  &  \\
\B &  &  &  &  &  &  \\
0 & 62.77 & 62.74 & - & 63.8 $\pm$ 0.6 & 61.37 & 72 \\
1 & 62.88 & 62.97 & - & 66.3 $\pm$ 0.8 & 55.53 & 72 \\
2 & 63.22 & 63.46 & - & 64.4 $\pm$ 2.0 & 58.14 & 73 \\
3 & 63.82 & 64.25 & - & 65.6 $\pm$ 3.0 & 59.74 & 75 \\
4 & 64.75 & 65.39 & - & 68.1 $\pm$ 4.0 & 59.6 & 76 \\
5 & 66.08 & 66.95 & - & 67.3 $\pm$ 5.0 & 60.79 & 78 \\
6 & 67.90 & - & - & - & 62.95 & 80 \\
7 & 70.28 & - & - & - & 64.45 & 82 \\
8 & 73.31 & - & - & - & 67.05 & 85 \\
9 & 77.04 & - & - & - & 70.41 & 88 \\
10 & 81.60 & - & - & - & 74.79 & - \\
11 & 86.75 & - & - & - & 80.11 & - \\
12 & 92.71 & - & - & - & 87.68 & - \\
13 & 99.70 & - & - & - & 91.68 & - \\ \bottomrule
\end{tabular}
\end{table*}

\subsubsection{Rotational Spectroscopy}
\alert{Our new trihybrid model contains the rotational transition data for all lower states  with energies less than 30,000 \cm{}, thus comprehensively incorporating all rotational hot bands; recall that the MoLLIST data had no rotational transitions. However, for applications within microwave astronomy that focus on detecting a small number of strong lines,  the existing CDMS data (collation \citep{CDMS1,CDMS2} sourced from \citet{CNrotdata1,CNrotdata2,CNrotdata3,CNrotdata4}) is likely preferable due to the incorporation of hyperfine splitting that cannot yet be included in a \Duo{} spectroscopic model (though future updates plan to add this feature to the program). CDMS has data for rotational transitions originating in the $v=0$ and $v=1$ states and can thus predict the strongest intensity hot bands.}

\alert{Averaging over hyperfine structure, our line positions and the CDMS values agree well with a RMSD of 0.002 \cm{}, while the agreement in the Einstein A coefficients is good with a RMSD of 0.012 s$^{-1}$, reflecting the close agreement between the experimental dipole moment used by CDMS and the equilibrium value of the \X{} state diagonal dipole moment curve used in our calculations.} %

\subsubsection{Lifetimes}

\alert{As discussed in \cite{21Mc.CN}, comparison of theoretical and experimental state lifetimes provide one of the most practical ways to validate the quality of \abinitio{} off-diagonal transition dipole moment curves and thus the predicted intensities for rovibronic transitions. Therefore, in \Cref{tab:lifetimes}, we compare our calculated excited state lifetimes with existing experimental and theoretical values.}

\alert{The \B{} state results in \Cref{tab:lifetimes} are straightforward; our results are within experimental uncertainties with strong agreement with \MOLLIST{} (i.e. \cite{14BrRaWe.CN}) and reasonable agreement with other theoretical results.  }

\alert{The \A{} state results, however, are more concerning. Our results are in close agreement with \MOLLIST{} and in reasonable agreement with other theoretical results. However, the two experimental results differ considerably (sometimes more than a factor of 2) from both each other and from all \abinitio{}-derived results. New experimental measurements would be highly desirable to validate or dispute the theoretical lifetimes (and thus the underlying dipole moment curves and predicted transition intensities). Astronomical or laboratory comparisons of the relative transition intensities in the \A-\X{} and \B-\X{} bands (perhaps near the 20,000 \cm{} region where both bands have comparable intensity) could also provide evidence to resolve this theory-experimental discrepancy.}

\section{Probe to constrain the variation of the proton-to-electron mass ratio}
CN was identified as a potentially promising molecular probe to constrain the variation of the proton-to-electron mass ratio in  \cite{19SyMoCu.CN}. This paper found that diatomic molecules with low lying electronic states show a sizeable amount of enhanced sensitive transitions to a variation of the proton-to-electron mass ratio. CN was further isolated as a potential probe due to its abundance and chemical properties. CN is one of very few molecules that has been detected at high redshift (z = 2.56 and 3.9) \citep{07RiWaCo.CN,07GuSaNe.CN} along with \ce{H2}, HCN, CO, \ce{HCO+}. The presence of CN at high red shift is significant when probing large time separation for constraining fundamental constants, such as the proton-to-electron mass ratio. A key driver of the construction of the spectroscopic model created in this work was to test the transitions of CN for their sensitivity and evaluate the efficacy of CN as a molecular probe.  

\subsection{Methodology}
In  \citet{19SyMoCu.CN,20SyMc.RN}, we simulated a variation in the proton-to-electron mass ratio in diatomic molecules using spectroscopic models. We calculated the sensitivity of the transitions, $K$, in these diatomic molecules by matching the transitions on their id number in each simulation and using equation \ref{eqn:mutrans} with the matched transition frequencies using
\begin{equation}
\centering
    \frac{\Delta \nu}{\nu} = K \frac{\Delta \mu}{\mu}, 
    \label{eqn:mutrans}
\end{equation}
where $\frac{\Delta \nu}{\nu} = \frac{\nu_{shifted} - \nu}{\nu}$ is the relative change in the transition frequencies, $K$ is the sensitivity coefficient, and $\frac{\Delta \mu}{\mu}$ is the fractional change in the proton-to-electron mass ratio, simulated by a shift in the molecular mass. 

As mentioned in the paper, there was a possibility that the id for the transitions was changed in the simulations and possible errors and miscalculations could arise. This method has been refined by matching the QNs of the energy levels from the states files and not using the transition id to identify a match. By matching on the QNs of the energy level we can account for the ordering of the states to change by the shift in the molecular mass. We have thus amended our approach to instead use
\begin{equation}
    K_{\mu}(i \rightarrow j) = \frac{E_jk_j - E_ik_i}{E_j - E_i},
    \label{eqn:K}
\end{equation}
where $E$ is the energy of the lower ($E_i$) and upper ($E_j$) energy levels of the transitions, and $k$ is the sensitivity coefficients of the lower ($k_i$) and upper ($k_j$) energy levels in a transition.

In the refinement of our method for calculating the sensitivity of transitions in diatomic molecules we found significant changes to the maximum sensitivity in some of the diatomic molecules investigated. While the conclusions we drew from our results in \cite{19SyMoCu.CN} remain the same, individual molecules have new results. The most dramatic change in the results for a single molecule was SiH, with a change in maximum $\Delta K$  from $\approx$9 to $\approx$900. 

We continue to use the terminology of `enhanced' transitions to describe transitions with a sensitivity coefficient $|K|>5$. Again we only consider transitions below the intensity cutoff of $I<10^{-30}$ cm$^{2}$ molecule$^{-1}$ at 1000 K.

\subsection{Results for CN}

\begin{figure}
    \centering
    \includegraphics[width = 0.5\textwidth]{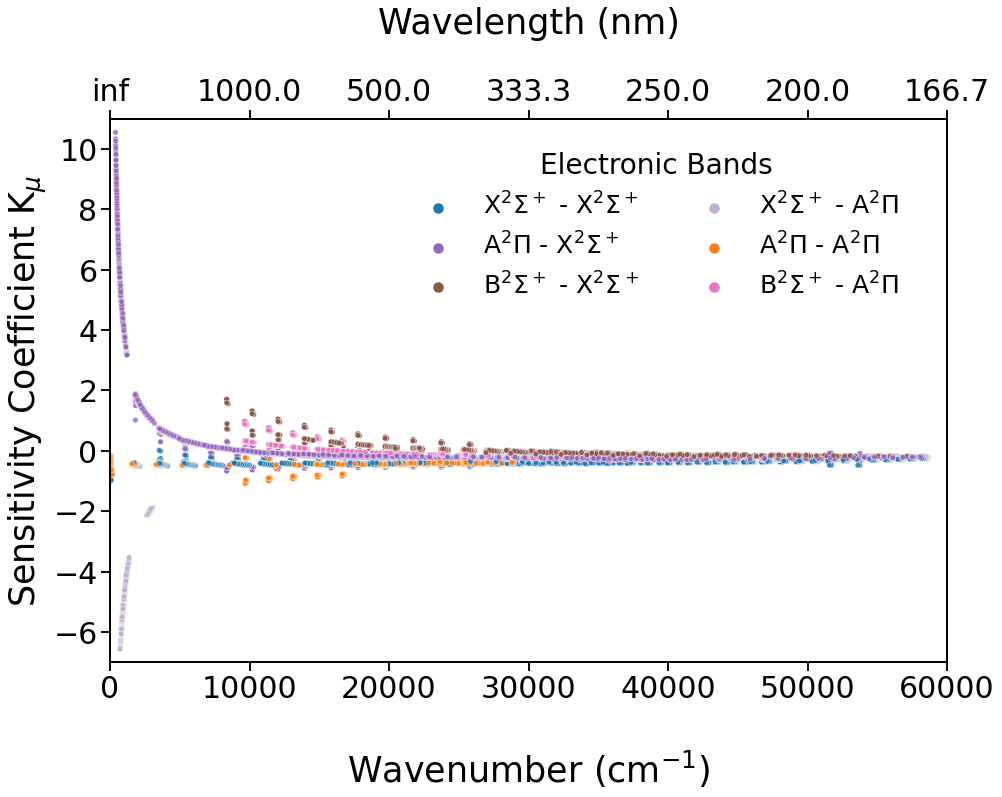}
    \caption{Spread of sensitivity of transitions within CN across energy.}
    \label{fig:mu}
\end{figure}

Similar to the results in \cite{19SyMoCu.CN} we see, in figure \ref{fig:mu}, that the enhanced transitions of CN have low transition frequencies (<1000 \cm) and involve the ground electronic state, \X{} and the first excited electronic state, \A. We note here that while there are 731 enhanced transitions in CN, all have intensities below $10^{-26}$ cm$^{2}$ molecule$^{-1}$ at 1000 K, and the lowest energy level of the enhanced transitions is greater than 9000 \cm. These limiting factors suggest that the enhanced transitions in CN might not be observable in the astrophysical environments that are probed to test the variation of the proton-to-electron mass ratio.

Given the observation of CN extra-galactically, it is worth considering whether stronger transitions with more modest $K$ could be useful. Restricting our search to intensities greater than $10^{-20}$ cm$^{2}$ molecule$^{-1}$ at 1000 K, we find that UV/optical transitions with $K$ values from -0.53 to 0.30 are observed, an order of magnitude larger than sensitivities obtained with \ce{H2} transitions. However, UV/optical transitions are yet to be observed extragalactically for CN. Rotational transitions that have been observed extragalactically all have the sensitivity of  pure rotational transitions, i.e. -1, not useful for a search for proton-to-electron mass variation. %

\subsection{Other diatomic molecules}
\begin{figure}
    \centering
    \includegraphics[width = 0.49\textwidth]{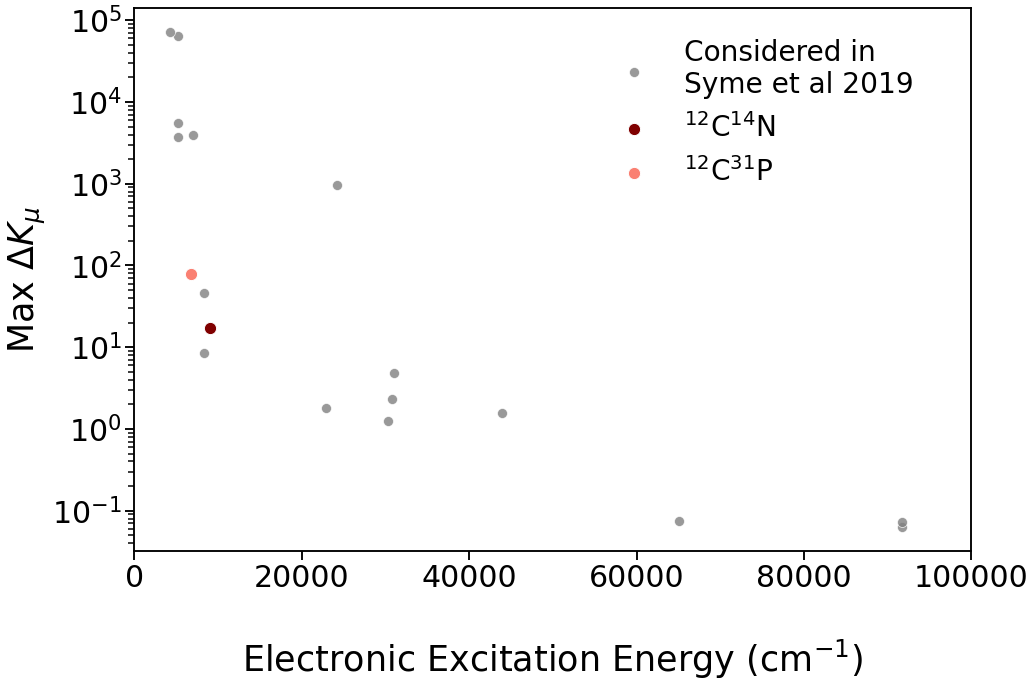}
    \caption{Relation between the energy difference of the ground electronic state and the first spin/symmetry allowed excited electronic state and the maximum $|\Delta K|$ with the new inclusions of CP, and CN.}
    \label{fig:DEDK}
\end{figure}

A key finding in \cite{19SyMoCu.CN} was the relationship between the energy of the first allowed excited state and the maximum sensitivity possible of the transitions considered. Here we provide an update on this relationship, shown in figure \ref{fig:DEDK} which has the addition of CN and the sensitivities calculated from the recently published spectroscopic model for CP \citep{21QuBaLi.CP}. We can see that while some of the data in this figure is different to that of figure 7 of \cite{19SyMoCu.CN} the relationship remains consistent.

\subsection{Future directions}
The refined method of calculating the sensitivity coefficients of transitions within molecules with the energy levels will allow us to scale up to larger molecules with reasonable computational cost.  We look forward to taking advantage of this with future work into the sensitivity to a variation of the proton-to-electron mass ratio of transitions within polyatomic molecules. %

\section{Conclusions} 
The complete \LLname{} line list for \ce{CN} can be found online at www.exomol.com and in the Supplementary Information in the ExoMol format. The line list contains 2,285,103 transitions between 28,004 energy levels from the 3 lowest electronic states (\X, \A, and \B) of the main isotopologue of CN. The final states file combines  energy levels from \Marvel, \Mollist, and ExoMol methodologies to produce a highly accurate and highly complete line list suitable for the full range of astrophysical applications from molecule detection using high-resolution cross-correlation techniques to modelling atmospheres to high precision. %

Of particular note is our method of constructing the states file that combines experimentally derived empirical energy levels, energy levels from model hamiltonians and from variationally determined energy levels to produce the most accurate states file. This is a good approach moving forward and we expect to see many future line lists utilising this approach of line list generation. %

A key motivator of this work was the development of a spectroscopic model to evaluate the efficacy of CN as a molecular probe to constrain the variation of the proton-to-electron mass ratio. While the sensitivities of the transitions within CN are not particularly enhanced, we do see some potential for CN to be used as a molecular probe  if UV/optical transitions can be detected as it is a common molecule in various astrophysical environments.

\section*{Acknowledgements} We would like to thank Juan C. Zapata Trujillo for insightful comments on the manuscript.

This research was undertaken with the assistance of resources from the National Computational Infrastructure (NCI Australia), an NCRIS enabled capability supported by the Australian Government.

The authors declare no conflicts of interest. 

\section*{Data Availability Statement}
The data underlying this article are available in the article and in its online supplementary material. These include the following files;
\begin{itemize}
    \item \Duo{} spectroscopic model and input file (12C-14N\_\_Trihybrid\_Duomodel.inp)
    \item \LLname{} states file (12C-14N\_\_Trihybrid.states);
    \item \LLname{} transitions file  for \ce{$^{12}$C$^14$N} (12C-14N\_\_Trihybrid.trans);
    \item Partitian function up to 10,000 K  for \ce{$^{12}$C$^14$N} (12C-14N\_\_Trihybrid.pf);
    \item Isotopologue states and transition files (in the isotopologue folder); 
    \item A sample ab initio input file (multi\_ci\_SO\_1.35.inp);
    \item csv file containing the PEC, TDM, and SOC from all of the \abinitio{} calculations (abinitio\_results.csv);
    \item Updated \Marvel{} transitions and energy levels (12C-14N\_MARVEL\_2021update.txt and 12C-14N\_MARVEL\_2021update.energies respectively). 
\end{itemize}
We note that the data generated in section 4 can be reproduced through the use of the CN spectroscopic model given here, as well as the available spectroscopic models on the ExoMol website (\url{www.exomol.com}), using the method described in the paper.

\section*{Additional Supporting Information}
\alert{As well as the data files described above, we include as supporting information two additional figures and associated discussion comparing (a) the new trihybrid and existing \Mollist{} line lists and (b) the new trihybrid line list created using the \Mollist{} X-X, A-X and B-X dipole moment curves compared to the the results when using our new smaller basis set result curves. }

For clarity, all supporting information files are described in README\_SI\_CN\_linelist.pdf.

\bibliography{CN_paper}
\bibliographystyle{mnras}

\end{document}